\pdfoutput=1

\documentclass[sigconf]{acmart}

\AtBeginDocument{%
  \providecommand\BibTeX{{%
    \normalfont B\kern-0.5em{\scshape i\kern-0.25em b}\kern-0.8em\TeX}}}
    
\copyrightyear{2022}
\acmYear{2022}
\setcopyright{rightsretained}
\acmConference[ASE '22]{37th IEEE/ACM International Conference on Automated Software Engineering}{October 10--14, 2022}{Rochester, MI, USA}
\acmBooktitle{37th IEEE/ACM International Conference on Automated Software Engineering (ASE '22), October 10--14, 2022, Rochester, MI, USA}
\acmDOI{10.1145/3551349.3561165}
\acmISBN{978-1-4503-9475-8/22/10}

\usepackage[inline]{enumitem}
\usepackage{pgfplots, pgfplotstable}
\setlist[enumerate,1]{label=\textit{\alph*)}}

\usepackage[utf8]{inputenc}

\title{Generalizability of Code Clone Detection on CodeBERT}

\author{Tim Sonnekalb}
\email{tim.sonnekalb@dlr.de}
\orcid{0000-0002-0067-1790}
\affiliation{%
	\institution{Institute of Data Sciene \\ German Aeorospace Center (DLR)}
	\city{Jena}
	\country{Germany}
}
\author{Bernd Gruner}
\email{bernd.gruner@dlr.de}
\orcid{0000-0002-4177-2993}
\affiliation{%
\institution{Institute of Data Sciene \\ German Aeorospace Center (DLR)}
\city{Jena}
\country{Germany}
}
\author{Clemens-Alexander Brust}
\email{clemens-alexander.brust@dlr.de}
\orcid{0000-0001-5419-1998}
\affiliation{%
\institution{Institute of Data Sciene \\ German Aeorospace Center (DLR)}
\city{Jena}
\country{Germany}
}
\author{Patrick Mäder}
\email{patrick.maeder@tu-ilmenau.de}
\orcid{0000-0001-6871-2707 }
\affiliation{%
\institution{Data-intensive Systems and Visualization Group  \\ Technical University Ilmenau}
\city{Ilmenau}
\country{Germany}
}

\begin{abstract}
	Transformer networks such as CodeBERT already achieve outstanding results for code clone detection in benchmark datasets, so one could assume that this task has already been solved. 
	However, code clone detection is not a trivial task. Semantic code clones, in particular, are challenging to detect.
	
	We show that the generalizability of CodeBERT decreases by evaluating two different subsets of Java code clones from BigCloneBench. We observe a significant drop in F1 score when we evaluate different code snippets and functionality IDs than those used for model building.
\end{abstract}



\date{\today}

\begin{document}

\maketitle

\section{Introduction}
Large language models, like transformers, are context-sensitive and should therefore be well suited for clone detection.
Microsoft's benchmark CodeXGlue \cite{codexglue2021} contains several machine learning for software engineering tasks solved by their CodeBERT model \cite{feng2020codebert}. One can observe the generalization capability of CodeBERT across tasks. One task is code clone detection which CodeXGlue models as binary classification. BigCloneBench (in Java) and POJ-104 (in C) are used as benchmark datasets.

BigCloneBench is a well-known code clone detection dataset that contains Java code snippets. Several methods \cite{guo2020graphcodebert, saini2018oreo, wang2020detecting, hammad2020deepclone} evaluate this benchmark dataset, making it a standard in this field. 
There are mainly 3 versions of BigCloneBench: BCB v1 \cite{Svajlenko2014BigCloneBench}, 
BCB v2 \cite{Svajlenko2016BigCloneEval} and CDLH (Clone Detection with Learning to Hash) \cite{wang2020detecting}. 
The latter is a pre-processed version of BCB v1 and is used in CodeXGlue \cite{codexglue2021}.



Code clone detection, in general, can be divided into four types of code clones. Type 1-3 are syntactic code clones with minor changes in their syntax. 
Type 4 clones or semantic clones 
are difficult to detect, as only a manual validation can safely check the different syntax but the same semantics. 
CodeBERT achieves an F1 score of 96.5\% for this task \cite{codexglue2021} which could be caused by group leakage. We verify this score by evaluating two different subsets of BigCloneBench not used for model building.


\section{Data} 
Svajlenko et al. \cite{Svajlenko2014BigCloneBench} create the BigCloneBench dataset (BCB v1) based on a sizable inter-project repository IJaDataset 2.0 of Java projects\footnote{The link is not available anymore, but Svajlenko et al. provide a mirror: https://github.com/clonebench/BigCloneBench}. 
They develop a semi-automated processing method to mine code clones based on ten given target similarity classes.
In a follow-up work  \cite{Svajlenko2016BigCloneEval}, the authors mine a larger version of BigCloneBench (BCB v2) with 43 code functionalities and provide the clone pairs in a database. 
 
The CDLH work \cite{wei2017supervised} presents another filtered version of BCB v1. The authors do not provide the pre-processed dataset or scripts. CodeXGlue uses the CDLH dataset and has published it with slightly different properties: instead of 9,134 given code snippets in CDLH, CodeXGlue contains 9,126. Though this dataset was filtered out of BCB v1, we reverse engineer 12 functionalities instead of 10. All properties are listed in Table \ref{tab:dataset-probs}.

After a detailed analysis of the dataset versions, we notice the following problems:
The datasets are highly imbalanced in their two classes, \textit{true clones} and \textit{false clones}, in their clone types, and in their functionalities. 

\begin{table}[ht]
	\centering
	\caption{Dataset properties in BigCloneBench versions.}
	\label{tab:dataset-probs}
	\begin{tabular}{ | l | r | r | r |}
		\hline
		\textbf{Property} & \textbf{BCB v1} & \textbf{BCB v2} & \textbf{CodeXGlue (CDLH)}  \\ 
		\hline
		Code snippets & 59,650 & 73,319 & 9,126 (9,134)\\ 
		Functionalities & 10 & 43 & 12 (10) \\
		Total clones & 6,508,632 & 8,863,185 & 1,731,857  \\ 
		True clones  & 6,246,167 & 8,584,153 & 1,170,336  \\
		False clones & 262,465 & 279,032 & 561,521 \\
		\hline
	\end{tabular}
\end{table}
\raggedbottom
%
%
%
The clone class distribution is due to the automatic creation process of the dataset. The true clones are assigned following an automated process, while the false clones are labeled manually by judges. For this reason, there are far fewer false clones than true ones (comp. Tab.~\ref{tab:dataset-probs}).

The clone types are determined by a similarity measure and are assigned to the Types \textit{Type-1}, \textit{Type-2}, \textit{Very Strongly Type-3}, \textit{Strongly Type-3}, \textit{Moderately Type-3}, and \textit{Weakly Type-3/Type 4} by hard-defined thresholds. Weakly Type-3 (syntactic) and Type-4 (semantic) clones cannot be separated with this approach. This class has a share of 98.2\% of the whole dataset (comp. with \cite{Svajlenko2014BigCloneBench}).

The authors \cite{Svajlenko2014BigCloneBench} create a search heuristic for the functionality classes. A blueprint for every functionality class is defined and compared with the code examples.
All 43 functionalities in BCB v2 are described here: \cite{hammad2020deepclone}. The functionality classes with the most examples have very general names like \textit{Copy} ($\sim$4.8M), \textit{Zip Files} ($\sim$966k), or \textit{Secure Hash} ($\sim$900k). We note that the full functionality of 25k software projects cannot be mapped to 43 classes in sufficient precision. Moreover, a code snippet can be assigned to multiple functionality classes if it matches multiple blueprints. BCB assigns only one functionality per example.
 
Another problem is that recall is only conditionally suitable as a measure for evaluating clone detection methods trained on previously labeled data. The original authors of BigCloneBench come to the same conclusion \cite{Svajlenko2015Evaluating}. Recall is the proportion of true positives out of all positives and cannot be computed unless all combinations of clone pairs are given. For a more significant number of examples, this cannot be checked for resource reasons, and therefore it cannot be ruled out that there are  code clones. An additional qualitative assessment on a small test subset can be helpful to obtain reliable measures.
 
\subsection{Evaluation Datasets}
CodeBERT uses only a small subset of $\sim$12\% of the entire BigCloneBench for model building. To test whether CodeBERT's generalization ability on the remaining code snippets is as high as the authors claim, we create two evaluation sets by filtering BCBv2:
\begin{itemize} [itemindent=-1em, topsep=5pt, parsep=2pt]
	\item [\textbf{1.}] \textbf{BCBv2 \textbackslash CDLH separated by clone IDs} We extracted a list with all code snippet IDs from training and validation sets of CDLH. From BCB v2, we removed all clone pairs with a matching id from this code ID list.
 	\item [\textbf{2.}] \textbf{BCBv2 \textbackslash CDLH separated func IDs} We created a list of clone pairs from the CDLH training and validation sets and reconstructed their functionality IDs. With this functionality ID list, we selected all clone pairs from BCB v2 not included in the CDLH functionality IDs.
\end{itemize}

The functionality IDs were not given in CodeXGlue, so we reverse engineer the functionality types of each example in the CDLH dataset. 
We succeed in recovering $\sim$60\% of the functionalities from the clone pairs. The clone IDs are independent of the functionalities. They have remained the same across versions of the dataset.

\section{Evaluation} 
We use the same hyperparameters and base model \textit{codebert-base} as the benchmark to ensure comparability.
We reproduce the fine-tuning of the model and run our evaluation on 4 Nvidia A100 GPUs.

The baseline of CodeXGlue was fine-tuned on only 10\% of training and 10\% of validation data \cite{codexglueGithub}. 
The train-valid-test fraction of CDLH is 52/24/24, which results in  $\sim$90k training and  $\sim$41k validation examples for fine-tuning. 
Table \ref{tab:eval-results} shows the evaluation results of our repeated fine-tuning of \textit{codebert-base} compared to the values given in the original papers and to our two created datasets.
The CodeXGlue paper does not provide precision and recall values. Our repetition yielded a slightly worse F1 score by 1.2\%. 


Evaluating our self-created subsets of BCBv2, we find a significant drop in F1 score of $\sim$27\% for dataset 1 and $\sim$47\% for dataset 2. Dataset 1 includes different clone IDs but partly code snippets with the same functionalities. Dataset 2 includes completely different functionalities, which could mean that the functionalities used for model building are too few to generalize across functionalities and other clone pairs.

Furthermore, transformer networks are limited by their input sequence length. CodeBERT's input limit is 800 tokens, which results in a sequence length of 400 per code snippet. More extended code snippets are truncated, which affects $\sim$32\% of the CodeXGlue code snippets. 

The bimodal CodeBERT model is pre-trained on the CodeSearchNet dataset, which includes pairs of documentation and code snippets from 6 programming languages, the percentage of Java in it is $\sim$24\%. Even if the fine-tuning was done for only one language, some tokens might have different cross-language semantics. Moreover, there is a gap in pre-training on combinations of natural language and programming languages and fine-tuning only on programming language pairs.

\def\arraystretch{1.2}%
\begin{table}
	\centering
	\caption{Evaluation metrics of CodeBERT in \%.}
	\label{tab:eval-results}
	\begin{tabular}{ | l | l | l | l | }
		\hline
		\textbf{Evaluation Dataset} & \textbf{Precision} & \textbf{Recall} & \textbf{F1 score} \\ 
		\hline
		\textbf{Lu et al. \cite{codexglue2021} (paper)} &  - & - & 95.6 \\ 
		\hline
		\textbf{Gou et al. \cite{guo2020graphcodebert} (paper)} &  94.7 & 93.4 & 94.1 \\
		\hline 
		\textbf{Lu et al. \cite{codexglue2021} (repeated)} &  93.03 & 95.82 & 94.40 \\
		\hline
		\textbf{\shortstack[l]{1. BCBv2 \textbackslash  CDLH \\ separated by clone IDs}} & 98.18 & 52.41 & 68.34 \\
		\hline
		\textbf{\shortstack[l]{2. BCBv2 \textbackslash  CDLH \\ separated by func IDs}} & 97.77 & 32.67 & 48.97 \\ 
		\hline
	\end{tabular}
\end{table}


\section{Conclusion}
We investigated the generalizability of the CodeBERT model by evaluating on unseen data from a larger portion of the same dataset. Although the results of CodeBERT for clone detection look good at first glance, there is still a significant drop in F1 score when evaluated on other code snippets and other functionalities.

We recommend the authors of CodeXGlue to revise their benchmark and use a more extensive training dataset by evaluating on the whole BigCloneBench dataset. Furthermore, BigCloneBench is a highly unbalanced dataset, so this must always be considered when pre-processing the data.

\newpage

\bibliographystyle{./ACM-Reference-Format}
\bibliography{./references}

\end{document}